\documentstyle[epsfig,preprint,aps]{revtex}

\def\gord{$ \raisebox{-.3ex}{$\stackrel{>}{_{\sim}}$} $}

\newcommand{\ba}{\begin{eqnarray}}
\newcommand{\ea}{\end{eqnarray}}
\newcommand{\be}{\begin{equation}}
\newcommand{\ee}{\end{equation}}
\newcommand{\bd}{\begin{displaymath}}
\newcommand{\ed}{\end{displaymath}}

\title{Thermal rates for baryon and anti-baryon production}

\author{Joseph Kapusta and Igor Shovkovy$^{*}$}
\address{School of Physics and Astronomy, University of Minnesota\\
Minneapolis, Minnesota 55455, USA}

\draft
\begin{document}

\maketitle

\begin{abstract}
We use a form of the fluctuation-dissipation theorem to derive formulas giving 
the rate of production of spin-1/2 baryons in terms of the fluctuations of 
either meson or quark fields.  The most general formulas do not assume 
thermal or chemical equilibrium.  When evaluated in a thermal ensemble we find 
equilibration times on the order of 10 fm/c near the critical temperature in 
QCD.
\end{abstract}
\vspace{0.25in}


\section{Introduction}

Stable or long-lived baryons are readily detectable in high energy nuclear
collisions, such as those at the CERN SPS (Super Proton Synchrotron) and
at the Brookhaven National Laboratory RHIC (Relativistic Heavy Ion
Collider).  They have masses on the order of 1 GeV, which makes them
relatively sensitive to the temperatures and expansion rates in such
collisions.  Recent measurements with Au+Au collisions at $\sqrt{s} = 200
A$ GeV at RHIC are consistent with protons, lambdas, sigmas, cascades, and
omegas, and their anti-particles, all being in chemical equilibrium at a
temperature of $170 \pm 10$ MeV \cite{QM2002}.  This temperature is close
to the expected critical or rapid crossover temperature in QCD between a
quark-gluon phase and a hadronic phase.  How is it possible to understand
such near perfect equilibration on the relatively short time scales of
high energy nuclear collisions?

One approach is kinetic theory, which was reviewed in some detail already
by Koch, M\"uller and Rafelski \cite{KMR}.  At finite temperatures, but
still in the hadronic phase, there exist many mesons, especially $\pi$,
$K$, $\rho$ and $\omega$.  Two body reactions like $\pi \pi \rightarrow B
\bar{B}$ dominate at moderate temperatures, say 100 MeV and less.  As the
temperature goes up so does the density of mesons.  Multi-particle
reactions, such as $\pi \pi \pi K \bar{K} \rightarrow B \bar{B}$ become
increasingly important.  These multi-particle reactions cannot be measured
directly in the laboratory, of course, but they can be related to the
inverse reactions by detailed balance.  Some of these inverse reactions,
those involving proton plus anti-proton annihilation into mesons, were
measured at LEAR (Low Energy Anti-proton Ring) at CERN.  Rapp and Shuryak
\cite{RS} have estimated that the sum of all reactions of the type $n \pi
\rightarrow p \bar{p}$ are able to yield fast equilibration times for
anti-protons at temperatures of order 150 to 170 MeV, perhaps as short as
several fm/c. However, there are at least three difficulties with applying
kinetic theory, using vacuum reactions rates, to high energy density
matter.  First, experimental data on the annihilation of hyperons into
mesons is practically nonexistent.  One must rely on approximate SU(3)
flavor symmetry to estimate the rates based on nucleon data.  Second,
there is no sound practical method to implement microscopic reactions
involving initial states with more than two particles in a microscopic
cascade computer code.  Two body initial state reactions are assumed to
occur when the colliding particles are within a distance
$\sqrt{\sigma/\pi}$ of each other, where $\sigma$ is the cross section.  
For three or more particles the criterion becomes ambiguous.  Third, and
perhaps most importantly, the very concept of localized interactions
occurring as in vacuum is no longer applicable.  When particle densities
reach 1 per cubic fermi with particles whose physical extent is also about
1 fm, it is not possible to define {\it in} and {\it out} states as in
the vacuum.  The interaction radius is comparable to the physical matter
radius of the hadrons.  The best that kinetic theory can do for hadrons at
high temperatures is to tell whether the equilibration time is small or
large but probably cannot give a quantitative number.

Another approach is with DCC (disoriented chiral condensates) \cite{DCC}.  
In this approach baryons are described as topological defects of the
chiral field using the model of Skyrme.  Domains are formed at some early
time, and as the matter expands all the domains must eventually line up
with the surrounding vacuum.  During this process defects are formed with
a probability that was first calculated in the context of the early
universe \cite{bigbang}.  The probability of defect production is
inversely proportional to the cube of the domain size.  Smaller domains
yield more baryons and anti-baryons.  Together with Wong, one of us showed
that the anomalously large number of $\Omega$ and $\bar{\Omega}$ observed
at the SPS could be understood in terms of this mechanism \cite{KW}.  The
typical domain size needed is 2 fm, which is just in the range predicted
by many different approaches to DCC formation \cite{size}.  However, it is
difficult to make much more quantitative calculations in this approach
without very extensive numerical simulations.  Even then, one may question
whether a low energy effective meson field theory can reasonably describe
variations over length scales as small as 2 fm.  Plus the relevant
frequencies involved are twice the proton mass.

At first it would seem that these two approaches are almost orthogonal to
each other.  In fact they are trying to describe the same physics, baryon
production at high energy density, starting from two opposite
perspectives.  Kinetic theory attempts to describe the process with many
mesons that propagate freely between localized collisions.  The DCC
approach assumes that the meson density is so high that they may be
treated collectively as a classical field; baryons arise as topological
defects of this field.

In this paper we attempt a more general description than either of the
above.  We use a version of the fluctuation-dissipation theorem, basically
the same formalism as used to compute the production rate of real and
virtual photons in hot matter.  The baryon production rate can be
expressed in terms of the fluctuations in the local meson fields or in
terms of fluctuations in the local quark fields.  The formulas derived can
be applied to systems in equilibrium or out of equilibrium.  In the former
case we evaluate the rates numerically; this is possible because of the
Boltzmann weighting of states.  In the latter case one must specify the
probability of different states according to the preparation of the
non-equilibrium system under consideration.  We first compute the rates
for non-strange baryons only, then we generalize to hyperons using SU(3)
flavor symmetry in the interaction Lagrangian.

\section{Nucleon Production}

In this section we analyze the production of nucleons and anti-nucleons by
vector and axial-vector fields or currents.  We first derive general
formulas that express the production rate in terms of fluctuations of
these fields or currents.  Formulas are given for arbitrary weighting of
states. The formulas are then evaluated in thermal equilibrium with a
Boltzmann weighting of states.  The vector and axial-vector fields or
currents are first expressed explicitly in terms of pion fields using the
nonlinear sigma model.  They are alternately expressed in terms of the
quark fields using vector meson dominance to determine the relevant
couplings.

\subsection{General formulas}

With a view toward the conventions used in the nonlinear sigma model we
write the coupling of nucleons to vector $V_{\mu}^a$ and axial-vector
$A_{\mu}^a$ currents or fields as
\be
{\cal L} = -\overline{\psi} \gamma^{\mu} \frac{\tau^a}{2} \psi V_{\mu}^a
+ g_A \overline{\psi} \gamma^{\mu} \gamma^5 \frac{\tau^a}{2} \psi A_{\mu}^a 
\, ,
\label{L-int}
\ee
where $g_A \approx 1.26$ is the axial coupling constant relative to the
vector.  We do not need to specify anything more about $V_{\mu}^a$ and
$A_{\mu}^a$.  They are given fields or currents that produce baryons via
the above coupling.  The production rate of nucleon/anti-nucleon pairs is
calculated in the same manner as dileptons \cite{dileps} using linear
response theory, equivalently a version of the fluctuation-dissipation
theorem.  We show explicitly the steps for production in the axial-vector
channel.

The matrix element for the process $i \rightarrow f +$ baryon $+$ 
anti-baryon is
\be 
S_{fi} = g_{A}
\langle f | \int d^{4} x A^{a}_{\mu}(x) J^{5\mu}_{a}(x) |i \rangle \, .
\ee
Here $J^{5\mu}_{a}$ is an abbreviation for the axial baryonic current as
expressed in the Lagrangian above.  The initial and final states are
arbitrary, usually fully interacting states except for the specific
interaction in Eq.~(\ref{L-int}).  The axial baryonic current,
corresponding to the baryons in the final state, is
\be
J^{5\mu}_{a}(x) = \frac{1}{V} \sqrt{\frac{m_N^2}{E_1 E_2}}
e^{-i x \cdot (p_1+p_2) } \bar{u}(p_1,s_1) 
\gamma^{5} \gamma^{\mu} \frac{\tau_{a}}{2} v(p_2,s_2) \, .
\ee
As a consequence of the translational invariance we can write 
\be
\langle f | A^{a}_{\mu}(x)|i \rangle 
= e^{i x \cdot k}
\langle f | A^{a}_{\mu}(0)|i \rangle \, .
\ee
Here $k\equiv k_{i}-k_{f}$ is the difference of the four-momenta of the
initial and final states.  The transition rate is
\be
R_{fi} =\frac{|S_{fi} |^2}{T V} \, , 
\ee
where $T$ is the time interval and $V$ is the volume.  Explicitly 
\ba
R_{fi} &=& g_{A}^{2}\int \frac{d^{4} x d^{4} x^{\prime}}{T V} 
\langle f | A^{a}_{\mu}(x+ x^{\prime}) J^{5\mu}_{a}(x+ x^{\prime}) 
|i \rangle \langle i | A^{b}_{\nu}(x^{\prime}) J^{5\nu}_{b}(x^{\prime})
|f\rangle   \nonumber \\
&=& \frac{g_{A}^{2}}{V^2} \frac{m_N^2}{E_1 E_2} \int d^{4} x
e^{ix\cdot (k-p_1-p_2)} \langle f | A^{a}_{\mu}(0) 
|i \rangle \langle i | A^{b}_{\nu}(0) |f \rangle \nonumber \\
&\times& \bar{u}(p_1,s_1)
\gamma^{5} \gamma^{\mu} \frac{\tau_{a}}{2} v(p_2,s_2)
\bar{v} (p_2,s_2) \gamma^{5} \gamma^{\nu}  \frac{\tau_{b}}{2}
u(p_1,s_1) \, .
\ea
Averaging over initial states with an arbitrary weight $w_i$ and summing
over final states, we arrive at the following differential rate.
\ba
dR &=&\frac{g_{A}^{2} m_N^2 }{E_1 E_2} 
\frac{d^{3}p_{1} d^{3}p_{2} }{(2\pi)^{6}} 
\sum_{i} w_i \sum_{f} 
(2\pi)^{4}\delta^{4}(p_1+p_2-k) \langle f | A^{a}_{\mu}(0)
|i \rangle \nonumber \\
&\times&  \langle i | A^{b}_{\nu}(0) |f \rangle
\mbox{Tr} \left( \frac{\not{p}_{1} +m_N}{2m_N} \gamma^{5} 
\gamma^{\mu} \frac{\tau_{a}}{2} \frac{\not{p}_{2}-m_N}{2m_N} 
\gamma^{5} \gamma^{\nu} \frac{\tau_{b}}{2}\right)
\ea
By introducing the correlation function
\be
A^{(-)ab}_{\mu\nu}(p) 
= - \sum_{i} w_i \sum_{f}
(2\pi)^{4}\delta^{4}(p-k) \langle f | A^{a}_{\mu}(0)|i \rangle 
\langle i | A^{b}_{\nu}(0) |f \rangle \, ,
\ee 
and evaluating the trace
\ba
&&\mbox{Tr} \left( \frac{\not{p}_{1} +m_N}{2m_N} \gamma^{5} 
\gamma^{\mu} \frac{\tau_{a}}{2} \frac{\not{p}_{2}-m_N}{2m_N}
\gamma^{5} \gamma^{\nu} \frac{\tau_{b}}{2}\right) \nonumber \\
&=& \frac{\delta^{ab}}{2m_N^2}\left[p_{1}^{\mu} p_{2}^{\nu}
+p_{1}^{\nu} p_{2}^{\mu} + g^{\mu\nu} \left(m_N^2
-p_{1} \cdot p_{2} \right)
\right] \, ,
\ea
we arrive at
\be
E_1 E_2\frac{dR_A}{d^{3}p_{1} d^{3}p_{2} }
=-\frac{g_{A}^{2}}{2 (2\pi)^{6}} 
A^{(-)aa}_{\mu\nu} (p_{1}+p_{2}) \left[p_{1}^{\mu} p_{2}^{\nu}
+p_{1}^{\nu} p_{2}^{\mu} - g^{\mu\nu} \left(p_{1} \cdot p_{2}
-m_N^2 \right)
\right] \, .
\ee
Apart from the rather trivial Lorentz tensor coming from the form of the
axial baryonic current, all the physics is contained in the correlation
function.

The corresponding rate arising from the vector interaction is easily
obtained.  The only differences are the replacement of the axial-vector
with the axial field or current, $A \rightarrow V$, setting $g_A^2 = 1$,
and changing the sign of the mass-squared term in the Lorentz tensor.
\be
E_1 E_2\frac{dR_V}{d^{3}p_{1} d^{3}p_{2} }
=-\frac{1}{2 (2\pi)^{6}} 
V^{(-)aa}_{\mu\nu} (p_{1}+p_{2}) \left[p_{1}^{\mu} p_{2}^{\nu}
+p_{1}^{\nu} p_{2}^{\mu} - g^{\mu\nu} \left(p_{1} \cdot p_{2}
+ m_N^2 \right)
\right]
\ee
These are the first significant results in this section.  They are quite
general, but they require knowledge of the correlation functions for the
initial states, appropriately weighted, and final states for the specific
system and conditions under consideration.

Now we evaluate the correlation functions in thermal equilibrium.  The
states $|i\rangle$ and $|f\rangle$ are conveniently assumed to be
eigenstates of the Hamiltonian $\hat{H}$ with eigenvalues $E_i$ and $E_f$.  
The weight is
\be
w_i = \frac{e^{-\beta E_i}}{Z} \, ,
\ee
where $\beta$ is the inverse temperature and $Z$ is the partition
function.  Following standard practice, the retarded correlation function
in position space is
\be
A^{(R)ab}_{\mu\nu}(x) = -i\frac{\theta(x^0)}{Z}\mbox{Tr} \left(
e^{-\beta \hat{H}} \left[A^{a}_{\mu}(x), 
A^{b}_{\nu}(0) \right] \right) \, .
\ee
Its Fourier transform is
\ba
&&A^{(R)ab}_{\mu\nu}(p)=-i\int_{0}^{\infty} d x^0
d^3 x e^{i\left[(p^0+i\varepsilon) x^0 -{\bf x} \cdot {\bf p}\right]}
\nonumber \\
&&\times \sum_{i,f}  
\left( w_i \langle i |  A^{a}_{\mu}(x) |f \rangle 
\langle f| A^{b}_{\nu}(0) |i \rangle 
- w_f \langle i | A^{a}_{\mu}(x) |f \rangle
\langle f| A^{b}_{\nu}(0) |i \rangle \right)\nonumber \\
&&= -i\int_{0}^{\infty} d x^0
d^3 x e^{i\left[(p^0-k^0+i\varepsilon) x^0 
-{\bf x} \cdot ({\bf p}-{\bf k})\right]}
\sum_{i,f} w_i \left(1-e^{\beta k^0}\right)\nonumber \\
&&\times
\langle i |  A^{a}_{\mu}(0) |f \rangle 
\langle f| A^{b}_{\nu}(0) |i \rangle
\nonumber \\
&&=\sum_{i,f} \frac{w_i\left(1-e^{\beta k^0}\right)
(2\pi)^{3} \delta^{3}({\bf p}-{\bf k}) }
{p^0-k^0+i\varepsilon}
\langle i |  A^{a}_{\mu}(0) |f \rangle 
\langle f | A^{b}_{\nu}(0) |i \rangle \, .
\ea
As before, $k = k_i - k_f$.  The imaginary part is
\be
\mbox{Im} A^{(R)ab}_{\mu\nu}(p) = -\frac{1}{2} 
\sum_{i,f} w_i \left(1-e^{\beta p^0}\right)
(2\pi)^{4} \delta^{4}(p-k)
\langle i |  A^{a}_{\mu}(0) |f \rangle
\langle f | A^{b}_{\nu}(0) |i \rangle \, .
\ee
Hence
\be
A^{(-)ab}_{\mu\nu}(p) = \frac{2}{1-e^{\beta p^0}}
\mbox{Im} A^{(R)ab}_{\mu\nu}(p) \, .
\ee
Exactly the same relationship holds in the vector channel with the
substitution of $A$ with $V$.

The sum of the rates of production by vector and axial-vector fields or
currents can now be written as
\ba
&& E_1 E_2 \frac{dR}{d^{3}p_{1} d^{3}p_{2}} 
=\frac{1}{(2\pi)^{6}} 
\frac{1}{e^{\beta (E_1 +E_2)}-1} \nonumber \\
&& \times \left\{ \mbox{Im} V^{(R)aa}_{\mu\nu}(p_1+p_2)
\left[p_{1}^{\mu} p_{2}^{\nu}
+p_{1}^{\nu} p_{2}^{\mu} - g^{\mu\nu} \left(
p_{1} \cdot p_{2} + m_N^2\right)\right] \right. \nonumber \\
&&\left. +g_A^2 \mbox{Im} A^{(R)aa}_{\mu\nu}(p_1+p_2)
\left[p_{1}^{\mu} p_{2}^{\nu}
+p_{1}^{\nu} p_{2}^{\mu} - g^{\mu\nu} \left(
p_{1} \cdot p_{2} - m_N^2\right)\right] \right\} \, .
\ea
This is the second significant result in this section.  In order to make
further progress, we need specific knowledge of the vector and
axial-vector fields or currents and their correlation functions.

\subsection{Production in terms of pion fluctuations}

The self-interactions among pions in the nonlinear sigma model are
contained in the Lagrangian 
\be
{\cal L} = \frac{f_{\pi}^{2}}{4}\mbox{Tr} \left(
\partial_{\mu} U \partial^{\mu} U^{\dagger} +
m_{\pi}^{2} U + m_{\pi}^{2} U^{\dagger} \right)
\ee
where 
\ba
U &=& \xi^2 \, ,\\
\xi &=& \exp\left(\frac{i}{2f_{\pi}}\pi^a \tau_a \right) \, .
\ea
Here $\pi^a$ is the pion field and $f_{\pi} \approx 93$ MeV is the pion
decay constant.  The pions couple to the baryons as in Eq.~(\ref{L-int})
with the derived currents
\ba
A^{a}_{\mu} &\equiv & \frac{i}{2}\mbox{Tr} \left[
\tau_{a} \left( \xi \partial_{\mu} \xi^{\dagger} 
-\xi^{\dagger} \partial_{\mu} \xi \right) \right] 
= \frac{1}{f_{\pi}} \partial_{\mu}\pi^{a} +\cdots \, , 
\label{V-def} \\
V^{a}_{\mu} &\equiv &
-\frac{i}{2}\mbox{Tr} \left[
\tau_{a} \left( \xi \partial_{\mu} \xi^{\dagger}
+\xi^{\dagger} \partial_{\mu} \xi \right) \right] 
= \frac{1}{f_{\pi}^{2}}
\varepsilon^{abc}\pi^{b}\partial_{\mu}\pi^{c}
+\cdots \, .
\label{A-def}
\ea
To first order in the pion field, this gives the usual derivative coupling
on account of the Goldberger-Treiman relation $g_A m_N = f_{\pi} g_{\pi
NN}$.  By expanding the interaction in powers of the pion field one may
derive the contribution to the production rate involving two, three, four,
five, etc.  number of pions in the initial state.

\subsection{Production in terms of quark fluctuations}

Since pions are made of quarks it should be possible to express the vector
and axial-vector fields or currents in terms of quark operators.  We can
make the connection using the hypothesis of vector meson dominance (VMD).  
In the two-flavor world the currents can be expressed in terms of the
$\rho$, $a_1$ and pion fields as
\ba
V_{\mu}^a &=& -g_{\rho NN} \rho_{\mu}^a \, , 
\label{Vmu-def}\\
A_{\mu}^a &=& \frac{g_{a_1NN}}{g_A} a_{1\mu}^a + \frac{g_{\pi NN}}{g_A m_N}
\partial_{\mu} \pi^a \, .
\label{Amu-def}
\ea
Using the Goldberger-Trieman relation and VMD the quark currents are
related to the hadronic fields.
\ba
\bar{q}\gamma_{\mu}\frac{\tau^a}{2}q &=& \frac{m_{\rho}^2}{g_{\rho\pi\pi}}
\rho_{\mu}^a  \\
\bar{q}\gamma_{\mu}\gamma^5 \frac{\tau^a}{2}q &=& 
\frac{m_{a_1}^2}{g_{a_1}} a_{1 \mu}^a + {\rm pion}
\ea
Apart from the pion pole we obtain the desired relationship.
\ba
V_{\mu}^a &=& - \frac{g_{\rho NN} g_{\rho \pi \pi}}{m_{\rho}^2}
\bar{q}\gamma_{\mu}\frac{\tau^a}{2}q \\
A_{\mu}^a &=& \frac{g_{a_1 NN} g_{a_1}}{m_{a_1}^2}
\bar{q}\gamma_{\mu} \gamma^5 \frac{\tau^a}{2}q
\ea
The imaginary part of the retarded correlator is given in terms of the
spectral density $\rho(s)$ as
\be
\mbox{Im} V^{(R)aa}_{\mu\nu}(k) = - 3 \left( g_{\mu\nu} -
\frac{k_{\mu} k_{\nu}}{k^2} \right) \pi \rho_V(s=k^2) \, ,
\ee
with a similar expression for the axial channel.  The factor of 3 arises
from the sum over isospin indices.

In our study of baryon/anti-baryon production, the spectral densities are
needed only for $s > 4m_N^2$. Then $s$ is large enough that they may be
computed using perturbative QCD. (The vector one may be measured directly
in electron-positron collisions; in the region of $\sqrt{s}$ of several
GeV a sum of hadronic resonances gives essentially the same answer, a nice
manifestation of duality.)  They are \cite{spectral}
\ba
\rho_V(s) &=& \left( \frac{g_{\rho NN} g_{\rho \pi \pi}}{m_{\rho}^2} \right)^2
\frac{s}{8\pi^2} \left( 1 + \frac{\alpha_s(s)}{\pi} + \cdots \right) \, , \\
\rho_A(s) &=& \left( \frac{g_{a_1 NN} g_{a_1}}{g_A m_{a_1}^2} \right)^2
\frac{s}{8\pi^2} \left( 1 + \frac{\alpha_s(s)}{\pi} + \cdots \right) \, .
\ea
Here $\alpha_s(s)$ is the QCD coupling evaluated at the scale $s$.

According to KSFR \cite{KSFR} $2 g_{\rho}^2 f_{\pi}^2 = m_{\rho}^2$.  
Furthermore it is usually assumed that $g_{\rho}$ is universal in the
sense that $g_{\rho NN} = g_{\rho \pi \pi}$, a result that holds rather
well numerically.  Then the coefficient of the vector spectral density is
especially simple.
\be
\left( \frac{g_{\rho NN} g_{\rho \pi \pi}}{m_{\rho}^2} \right)^2
= \frac{1}{4f_{\pi}^4}
\ee
In the absence of any better information it is quite reasonable to assume
that the same holds in the axial-vector channel.
\be
\left( \frac{g_{a_1 NN} g_{a_1}}{ m_{a_1}^2} \right)^2
= \frac{1}{4f_{\pi}^4}
\ee
This means that the vector and axial-vector currents have the same
spectral density since the axial coupling $g_A$ then cancels out.  Equal
contributions is quite natural when the up and down quark masses are very
small, as is the case in the real world.

The tensor algebra can now be done using $k = p_1 + p_2$.  It yields the
third significant result of this section.
\be
E_1 E_2 \frac{dR}{d^{3}p_{1} d^{3}p_{2}} = \frac{3}{8(2\pi)^7} 
\frac{1}{e^{\beta (E_1+E_2)}-1} \frac{s(s-m_N^2)}{f_{\pi}^4}
\left( 1 + \frac{\alpha_s(s)}{\pi} + \cdots \right)
\ee
This result is quite remarkable in that the rate is inversely proportional
to the fourth power of $f_{\pi}$ and does not depend on any other hadronic
parameters except the nucleon mass.

Finally, the overall rate of production may be computed by integrating
over the momenta of the outgoing nucleons.  Since the threshold energy,
$2m_N$, is much greater than the temperatures envisioned, $T < 200$ MeV,
it suffices to drop the minus one in the Bose-Einstein distribution factor
in the differential rate.  Then, with $K_n$ denoting the Bessel function
of the second kind, we get
\ba
R(\bar{N}N) &=& \frac{9}{(2\pi)^5} \left(
1+\frac{\alpha_{s}(4m_N^2)}{\pi} + \cdots \right) 
\frac{m_N^{4} T^{4}}{f_{\pi}^{4}} 
\left[\frac{m_N^{2}}{T^{2}}
K_{1}^{2}\left(\frac{m_N}{T}\right)
\right. \nonumber \\
&+& \left.
4 \frac{m_N}{T} K_{1}\left(\frac{m_N}{T}\right)
K_{2}\left(\frac{m_N}{T}\right)
+\left(8+\frac{m_N^{2}}{T^{2}}\right)K_{2}^{2}
\left(\frac{m_N}{T}\right)\right] \, .
\ea
In the nonrelativistic limit this becomes
\be
R(\bar{N}N) = \frac{9}{2(2\pi)^4} \left(
1+\frac{\alpha_{s}(4m_N^2)}{\pi} + \cdots \right) 
\frac{m_N^5 T^3}{f_{\pi}^{4}}
\exp\left(-2m_N/T\right) \, .
\ee
This is the total rate for the production of $\bar{p}p$, $\bar{p}n$,
$\bar{n}p$ and $\bar{n}n$.  The individual rates are related as:
$R(\bar{p}n) = R(\bar{n}p) = 2R(\bar{p}p) = 2R(\bar{n}n)$.

It should be noted that we have evaluated $\alpha_s$ at threshold for
nucleon/anti-nucleon production where the rate is a maximum.  According to
the latest analysis \cite{PDG} $\alpha_s(m_{\tau}^2) = 0.35 \pm 0.03$.  
Since $m_{\tau} = 1777$ MeV is very close to $2m_N$ we use that as the
numerical value in our later calculations.  The first perturbative
correction to the spectral density is only about 10\%.

It should also be noted that we have included the interactions involving
the isospin $I = 1$ currents only.  If the isospin $I = 0$ currents were
included too one might expect the rates for $\bar{p}p$ and $\bar{n}n$ to
increase while the rates for $\bar{p}n$ and $\bar{n}p$ to be unchanged.  
Indeed, this is what happens, and in fact the rates for $\bar{p}p$ and
$\bar{p}n$ become approximately equal, but we defer the actual analysis to
the next section.

Finally, we point out that since nucleons are composite objects they have
form factors.  These form factors will multiply the above rates and will
serve to decrease them to some degree.  We will defer the determination of
the form factors to the next chapter.  The reason is that there is
accurate data on $\bar{p}p$ annihilations, and to do a precise analysis we
must first include the isospin $I=0$ current.

\section{Nucleon and Hyperon Production}

An amazing fact in heavy ion collisions at RHIC is that hyperons are
produced in relatively great abundance.  Measurements indicate that they
are, for all practical purposes, in chemical equilibrium at a temperature
of $170 \pm 10$ MeV.  Therefore it behooves for us to analyze hyperon
production.  In addition, we now will include coupling to the isospin $I =
0$ vector and axial-vector currents or fields too.  We side-stepped that
contribution in the last section for clarity of presentation and to avoid
making phenomenological estimates of the relevant couplings: assumption of
SU(3) invariance of the interactions will help to resolve that issue.  We
will consider fluctuations in the SU(2) sector only as well as
fluctuations in the full SU(3) sector.  The results are rather different.

\subsection{General SU(3) invariant couplings}

Flavor SU(3) is not nearly as good a symmetry as SU(2).  However,
experience over many decades of research has shown that it is usually a
very good reproduction of experimental data to put all the flavor symmetry
breaking in the mass terms but to insist that the interactions be flavor
symmetric.  That is what we shall do too.

The vector meson nonet is conventionally written as follows \cite{G,CL}.
\be
{\cal V} = \left(\begin{array}{ccc}
\frac{\rho^0}{\sqrt{2}}+\frac{\omega_8}{\sqrt{6}} +\frac{\omega_s}{\sqrt{3}} 
&\rho^+ & K^{*+}\\
\rho^- & -\frac{\rho^0}{\sqrt{2}}+\frac{\omega_8}{\sqrt{6}} 
+\frac{\omega_s}{\sqrt{3}} & K^{*0}\\
K^{*-} & \bar{K}^{*0}& \frac{-2\omega_8}{\sqrt{6}}+\frac{\omega_s}{\sqrt{3}}\\
\end{array}\right)
\ee
The singlet and octet components are actually a mixture of the physical
$\omega$ and $\phi$ mesons.
\ba
\omega_8 &=& \phi \cos\theta_V + \omega \sin\theta_V \\
\omega_s &=& \omega \cos\theta_V - \phi \sin\theta_V
\ea
Ideal mixing occurs when the $\omega$ has no $s\bar{s}$ component while
the $\phi$ is pure $s\bar{s}$.  The ideal mixing angle is $\tan\theta_{\rm
ideal} = 1/\sqrt{2}$, or $\theta_{\rm ideal} \approx 35.3^o$.  
Experimentally the mixing angle seems to be about $39^o$ \cite{PDG}.  We
will approximate the mixing as ideal to simplify formulas.  Such fine
details are not likely to be important in the context we have in mind,
namely, heavy ion collisions.  Therefore we use the nonet representation.
\be
{\cal V} = \left(\begin{array}{ccc}
\frac{\rho^0}{\sqrt{2}}+\frac{\omega}{\sqrt{2}} &\rho^+ & K^{*+}\\
\rho^- & -\frac{\rho^0}{\sqrt{2}}+\frac{\omega}{\sqrt{2}} & K^{*0}\\
K^{*-} & \bar{K}^{*0}& -\phi \\
\end{array}\right)
\ee
A similar assumption of ideal mixing in the axial-vector meson nonet
yields the following representation.
\be
{\cal A} = \left(\begin{array}{ccc}
\frac{a_{1}^0}{\sqrt{2}}+\frac{f_{1}(1285)}{\sqrt{2}} &a_{1}^+ & K_{1}^{+}\\
a_{1}^- & -\frac{a_{1}^0}{\sqrt{2}}+\frac{f_{1}(1285)}{\sqrt{2}} & K_{1}^{0} \\
K_{1}^{-} & \bar{K}_{1}^{0} & -f_1(1420)\\
\end{array}\right)
\ee
Finally we give the matrix of the baryon octet.
\be
{\cal B} = \left(\begin{array}{ccc}
\frac{\Sigma^0}{\sqrt{2}}+\frac{\Lambda^0}{\sqrt{6}} &\Sigma^+ & p\\
\Sigma^- & -\frac{\Sigma^0}{\sqrt{2}}+\frac{\Lambda^0}{\sqrt{6}} & n \\
\Xi^- & \Xi^0 & \frac{-2\Lambda^0}{\sqrt{6}} \\
\end{array}\right)
\ee

As is well known there are three types of SU(3) invariant couplings:  the
F and D types, so-called because they involve the correspondingly labeled
group structure constants, and the singlet coupling, which involves the
trace of the meson matrix.
\ba
{\cal L}_{\rm int} & = & \frac{g_{\rho NN}}{\sqrt{2}}\Bigg[
(1-\alpha_V) {\rm Tr}\left( \bar{{\cal B}} \gamma^{\mu}
\left[ {\cal V}_{\mu},{\cal B} \right] \right) 
+\beta_V{\rm Tr}\left( \bar{{\cal B}} \gamma^{\mu} {\cal B} \right)
 {\rm Tr}\left( {\cal V}_{\mu} \right) \nonumber \\
&+& \alpha_V {\rm Tr}\left( \bar{{\cal B}} \gamma^{\mu}
\left\{ {\cal V}_{\mu},{\cal B} \right\} \right)
+g_{A}\alpha_A {\rm Tr}\left( \bar{{\cal B}} \gamma^{\mu} \gamma^5
\left\{ {\cal A}_{\mu},{\cal B} \right\} \right) \nonumber \\
&+& g_{A}(1-\alpha_A) {\rm Tr}\left( \bar{{\cal B}} \gamma^{\mu} \gamma^5
\left[ {\cal A}_{\mu},{\cal B} \right] \right) 
+ g_{A} \beta_A {\rm Tr}\left( \bar{{\cal B}} 
\gamma^{\mu} \gamma^5 {\cal B} \right)
 {\rm Tr}\left( {\cal A}_{\mu} \right)\Bigg]
\ea
The overall normalization of this interaction Lagrangian is determined by
the coupling of the nucleons; see Eqs.~(\ref{L-int}) and (\ref{Vmu-def}).  
Four parameters are introduced: $\alpha_V$ and $\alpha_A$, which determine
the relative contributions of the D and F type couplings in the vector and
axial-vector channels, respectively, and $\beta_V$ and $\beta_A$, which
determine the corresponding singlet contributions.

There was evidence already in the 1960's that $\alpha_{A}$ was about $2/3$
\cite{G}. This has been confirmed repeatedly over the years.  For example,
in their analysis of the spin content of the nucleon Close and Roberts
\cite{CR} determined that $\alpha \approx 0.635$. As another example,
Klingl, Kaiser and Weise \cite{KKW} use vector meson dominance together
with SU(3) symmetry to deduce $\alpha_A = 0.68$.  We shall therefore fix
$\alpha_A = 2/3$.

The value of $\alpha_{V}$ is determined by the requirement that the
coupling of the $\phi$ vector meson, which has already been taken to be a
pure $s\bar{s}$ state, to nucleons vanishes: $g_{\phi NN}=0$ \cite{KKW}.
This is just one aspect of the OZI rule \cite{OZI}.  This requirement
fixes $\alpha_{V}=(1-\beta_V)/2$.

We enforce the standard ratio of coupling constants of $\omega$ and $\rho$
vector mesons to nucleons, $g_{\omega NN} = 3 g_{\rho NN}$, as follows
from the quark model and the conventional definition of $\rho$ and
$\omega$ currents, see Eqs.~(\ref{def-rho}) and (\ref{def-omega}).  Thus,
we determine $\beta_{V}=1$. This further implies that $\alpha_V=0$.  
Finally, we require that the coupling of the nucleon to the $f_{1}(1420)$
meson vanishes, in analogy to the vanishing coupling of the $\phi$ meson
to the nucleon.  This condition fixes $\beta_{A}=(1-2\alpha_{A}) = -1/3$.

The relative couplings in the vector channel with the choice
$\alpha_{V}=0$ and $\beta_{V}=1$ are shown in Table I. The absolute
normalization may be inferred from the nucleon-nucleon couplings.  The
corresponding couplings in the axial-vector channel with the choice
$\alpha_A = 2/3$ and $\beta_A = -1/3$ are shown in Table II.

\subsection{Rates}

The invariant differential rates for all baryons in the octet can now be
inferred.  The only missing pieces are the spectral densities in the
various channels.  The currents as conventionally defined are, for
example,
\ba
j_{\rho^0}^{\mu}&=&\frac{1}{2}(\bar{u}\gamma^{\mu}u-\bar{d}\gamma^{\mu}d)
\label{def-rho} \\
j_{\omega}^{\mu}&=&\frac{1}{6}(\bar{u}\gamma^{\mu}u+\bar{d}\gamma^{\mu}d)
\label{def-omega}\\
j_{\phi}^{\mu}&=&-\frac{1}{3}\bar{s}\gamma^{\mu}s
\label{def-phi}
\ea
and so on. The spectral densities for these currents from perturbative 
QCD are \cite{spectral}
\ba
\rho_{\rho^0}(s)&=&\frac{s}{8\pi^2}\left(1+\frac{\alpha_s(s)}{\pi}\right)
\\
\rho_{\omega}(s)&=&\frac{s}{72\pi^2}\left(1+\frac{\alpha_s(s)}{\pi}\right)
\\
\rho_{\phi}(s)&=&\frac{s}{36\pi^2}\left(1+\frac{\alpha_s(s)}{\pi}\right)
\, .
\ea
These are applicable above some threshold value which is always above the
threshold for production of the corresponding baryon/anti-baryon pair,
typically 1.5 to 2.5 GeV$^2$. These threshold values could be estimated
rather well by making use of the QCD sum rules \cite{spectral}.

The rates are now determined from the SU(3) couplings, given in Tables I
and II, and the spectral densities, given above.  The relative weights are
given in Tables III and IV from which the rates may be inferred.  For
convenience define the pair of functions
\be
r_{\pm}(m_1,m_2) = \frac{2}{(4\pi)^7} 
\frac{F_{\rm ANN}^2(s)}{e^{\beta (E_1+E_2)}-1} 
\frac{2s^2-(m_1^2+m_2^2)s-(m_1^2-m_2^2)^2 \pm 6m_1m_2s}{f_{\pi}^4}
\left( 1 + \frac{\alpha_s}{\pi} \right) 
\ee 
where the $\pm$ corresponds to vector/axial-vector contributions.  The
function $F_{\rm ANN}(s)$ is a form factor, alluded to in the previous
section, and determined in the following subsection.  Using the symbol $r$
as shorthand notation for $E_1E_2dR/d^3p_1d^3p_2$ some examples are given
below.
\ba
r(n\overline{p}) &=& 2 r_+(m_N,m_N) + 2 r_-(m_N,m_N) 
\label{npbar}\\
r(p\overline{p}) &=& 2 r_+(m_N,m_N) + \frac{82}{81} r_-(m_N,m_N) 
\label{ppbar}\\
r(\Lambda\overline{p}) &=& 3 r_+(m_{\Lambda},m_N) + 
\frac{25}{27} r_-(m_{\Lambda},m_N) \\
r(\Xi^-\overline{\Lambda}) &=& 3 r_+(m_{\Lambda},m_{\Xi}) + 
\frac{1}{27} r_-(m_{\Lambda},m_{\Xi}) 
\ea
Altogether there are 46 nonvanishing combinations of baryon/anti-baryon
pairs.

The rates cannot in general be evaluated in closed form with the form
factor included.  However, it turns out that a very good approximation
(within 10\% at $T=200$ MeV) is to evaluate $F_{\rm ANN}^2(s)$ at the
average value $\bar{s} = (m_1+m_2)^2 + 3(m_1+m_2)T$, as discussed in the
next subsection.  Let us define
\be
R_{\pm} = \int \frac{d^3p_1}{E_1} \frac{d^3p_2}{E_2} r_{\pm} \, .
\ee
If we can evaluate $F_{\rm ANN}^2$ at the average value of $s$ then the
integral can be done in closed form.  It is
\ba
R_{\pm} &=& \frac{9(1+\alpha_s/\pi)T^8}{4(2\pi)^5f_{\pi}^4} z_1^2 z_2^2
\left\{ 4z_1K_1(z_1)K_2(z_2) + 4z_2K_1(z_2)K_2(z_1) \right. \nonumber \\
&& \left. \pm (z_1 \pm z_2)^2 K_1(z_1)K_1(z_2) +
[16+(z_1\pm z_2)^2] K_2(z_1)K_2(z_2) \right\}
F_{\rm ANN}^2(\bar{s}) \, ,
\ea
where $z_i=m_i/T$.

\subsection{Form factors}

Nucleons are composite objects, hence they have form factors that depend
on the specific process.  The appropriate form factor here is not the
electric or magnetic form factor since we are coupling the nucleons to
mesonic currents, ultimately expressed in terms of the quark fields.  We
can obtain a very good estimate of the relevant form factor by comparing
the rate for $\bar{p}p$ production, as derived above, with the rate for
$\bar{p}p$ annihilation, as obtained from kinetic theory.

The kinetic theory expression for the annihilation rate is
\be
E_1 E_2 \frac{dR_{\rm ANN}}{d^{3}p_{1} d^{3}p_{2}}(\bar{p}p) = 
\frac{4}{(2\pi)^6} f(E_1) f(E_2)  
\frac{\sqrt{(p_1 \cdot p_2)^2-m_N^4}}{E_1E_2}
\sigma_{\rm ANN}^{\bar{p}p}(s) \, ,
\ee
where $\sigma_{\rm ANN}^{\bar{p}p}(s)$ is the annihilation cross section
which specifically excludes a baryon/anti-baryon pair in the final state.  
If we approximate the thermal distributions $f(E) = \exp(-E/T)$, which is
a very good approximation to the Fermi-Dirac distribution at the modest
temperatures of relevance here, we obtain the following simple expression.
\be
E_1 E_2 \frac{dR_{\rm ANN}}{d^{3}p_{1} d^{3}p_{2}}(\bar{p}p) = 
\frac{2}{(2\pi)^6} \exp(-(E_1+E_2)/T)  
\sqrt{s(s-4m_N^2)} \sigma_{\rm ANN}(s)
\ee
In chemical equilibrium the rate for production must be equal to the rate
for annihilation.  The former is given in Eq.~(\ref{ppbar}).  Dropping the
$f_1$ contribution for simplicity of presentation, meaning that the factor
82/81 is set to 1, which is an approximation to the rate better than
$0.5\%$, we get
\be
E_1 E_2 \frac{dR}{d^{3}p_{1} d^{3}p_{2}}(\bar{p}p) = 
\frac{3}{32(2\pi)^7} \exp(-(E_1+E_2)/T)  
\frac{s^2}{f_{\pi}^4} \left( 1 + \frac{\alpha_s}{\pi}\right)
F_{\rm ANN}^2(s) \, .
\ee
Equating these two yields an expression for the form factor in terms of
the annihilation cross section.
\be
F_{\rm ANN}^2(s) = \frac{128\pi}{3} \frac{f_{\pi}^4}{1+\alpha_s/\pi} \,
\sqrt{1-\frac{4m_N^2}{s}} \, \frac{\sigma_{\rm ANN}^{\bar{p}p}(s)}{s}
\ee
The experimentally measured values of the $\bar{p}p$ annihilation cross
section from several hundred MeV/c to 8 GeV/c lab momentum has been
meticulously parameterized by Cugnon and Vandermeulen \cite{Cugnon}; see
also the review by Dover, Gutsche, Maruyama and Faessler \cite{Dover}.  
The fit is
\be
\sigma_{\rm ANN}^{\bar{p}p}(p_L)=\frac{38}{\sqrt{p_L}}+\frac{24}{p_L^{1.1}}
\ee
given in mb when the lab momentum $p_L$ is given in GeV/c.

We have fit the form factor with the monopole function
\be
F_{\rm ANN}(s) = \frac{1}{2.21 + (s-4m_N^2)/\Lambda^2} \, ,
\ee
where $\Lambda = 1.63$ GeV.  This function gives a very good
representation for $s-4m_N^2 > 0.5$ GeV$^2$, but overestimates $F_{\rm
ANN}$ by about 10\% at $s=4m_N^2$. This overestimate is acceptable because,
as we shall see below, the average value of $s-4m_N^2$ is greater than 
$0.5$ GeV$^2$ for $T> 100$ MeV.

For hyperons we choose the parameterization of the form factor to be
\be
F_{\rm ANN}(s) = \frac{1}{2.21 + [s-(m_1+m_2)^2]/\Lambda^2} \, ,
\ee
with the same value of $\Lambda$.

The total and elastic cross sections for $\bar{p}n$ have been measured,
and the annihilation cross section estimated, for lab kinetic energies
between 450 and 1068 MeV \cite{Elioff} and at 3.5 GeV \cite{Reynolds}.  
The annihilation cross section for $\bar{n}p$ has been explicitly measured
for lab momenta between 100 and 500 MeV/c \cite{Armstrong}.  In all these
cases the annihilation cross section for $\bar{p}n$ and $\bar{n}p$ has
been equal to the annihilation cross section for $\bar{p}p$, albeit with
large error bars in the two former cases.  The difference in the thermal
production rates between $\bar{p}p$ and $\bar{n}p$ calculated here is well
within the error bars. [Note that the rates in Eqs.~(\ref{npbar}) and 
(\ref{ppbar}) do not differ much because $r_{-}(m_{N},m_{N})$ is rather 
small compared to $r_{+}(m_{N},m_{N})$ for $s\gord 4m_{N}^{2}$.] This is
a gratifying conclusion which points to the consistency of our results 
compared to experimental data.

In principle the differential rates must be integrated over all energies,
including the $s$ dependence of the form factor.  A rough approximation is
to evaluate the form factor at the thermal average value of $s$, which is
$\bar{s} = (m_1+m_2)^2+3(m_1+m_2)T$, in the nonrelativistic limit and
dropping terms of relative order $(T/m)^2$.  It turns out that this
approximation is good to better than 10\% for temperatures less than 200
MeV.

\section{Numerical Results}
   
Consider a system at fixed temperature and volume but not necessarily in
chemical equilibrium with respect to the baryons.  If one of the baryons,
say the anti-proton for definiteness, is out of chemical equilibrium for
any reason, how long does it take for it to return to equilibrium?  The
rate equation for the density is
\be
\frac{dn_{\bar{p}}}{dt} = \sum_{b} R(b\bar{p})
\left[ 1 - \frac{n_{\bar{p}} n_b}{n_{\bar{p}}^{\rm equil} n_b^{\rm equil}}
\right] \, ,
\ee
where $n_{\bar{p}}^{\rm equil}$ is the equilibrium density for
anti-protons and $n_b^{\rm equil}$ is the equilibrium density for the
baryon species $b$.  The characteristic time scale for bringing the
anti-protons to equilibrium is
\be
\tau_{\bar{p}} = n_{\bar{p}}^{\rm equil}/\sum_b R(b\bar{p}) \, .
\ee
This characteristic time is more intuitive than the rates themselves.  A
fully dynamical model of the evolution of matter is required for the
detailed knowledge of how the abundances develop, but for the purpose of
gaining insight to the dynamics the characteristic equilibration time is
perhaps more useful.
  
\begin{figure}
\epsfig{file=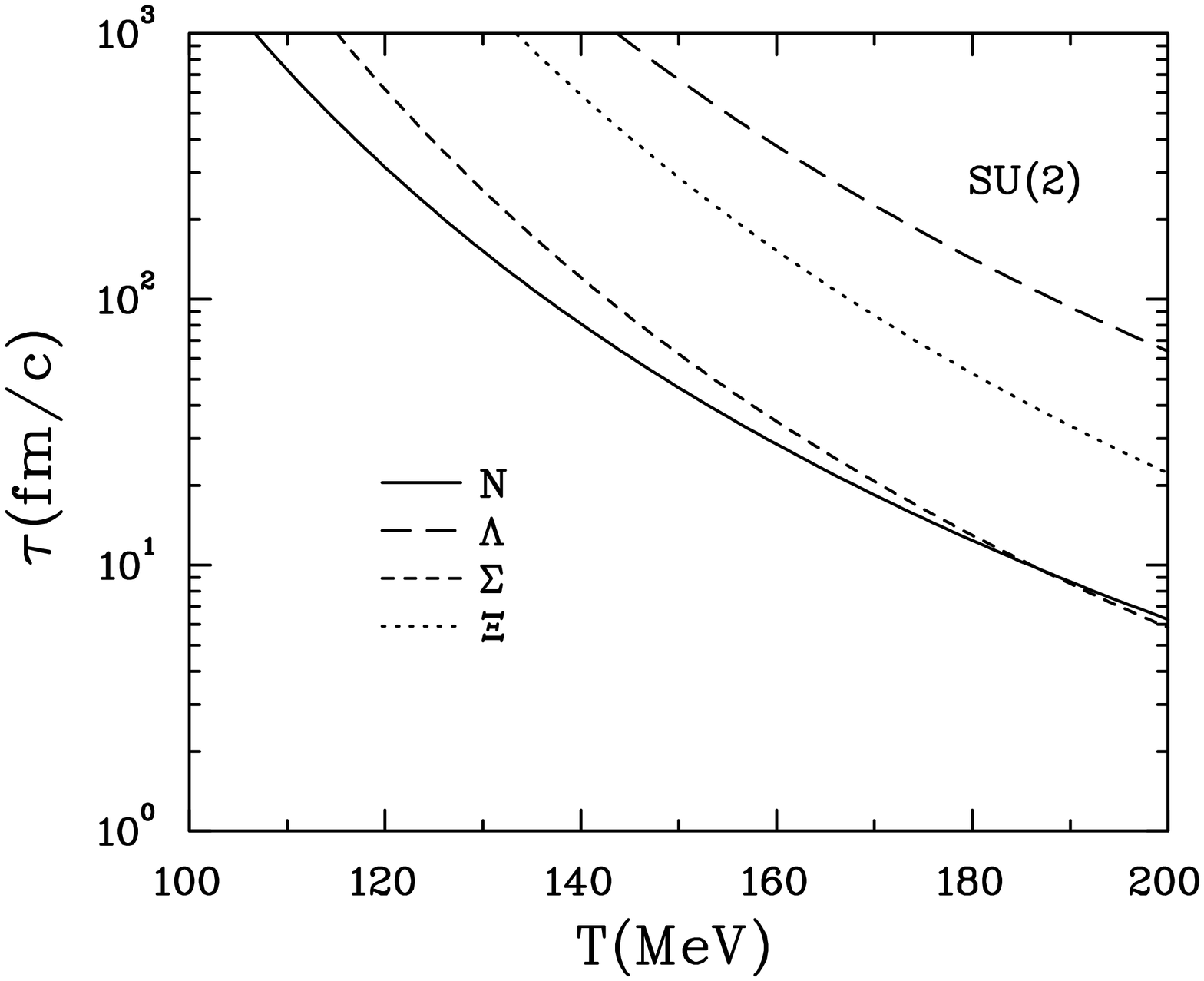,width=14cm,angle=90,bbllx=4pt,bblly=62pt,bburx=564pt,bbury=666pt}
\caption{Numerical results for equilibration times in the
case when only the fluctuations in the $SU(2)$ meson sector 
are taken into accout.}
\label{fig_tau_su2}
\end{figure}

\begin{figure}
\epsfig{file=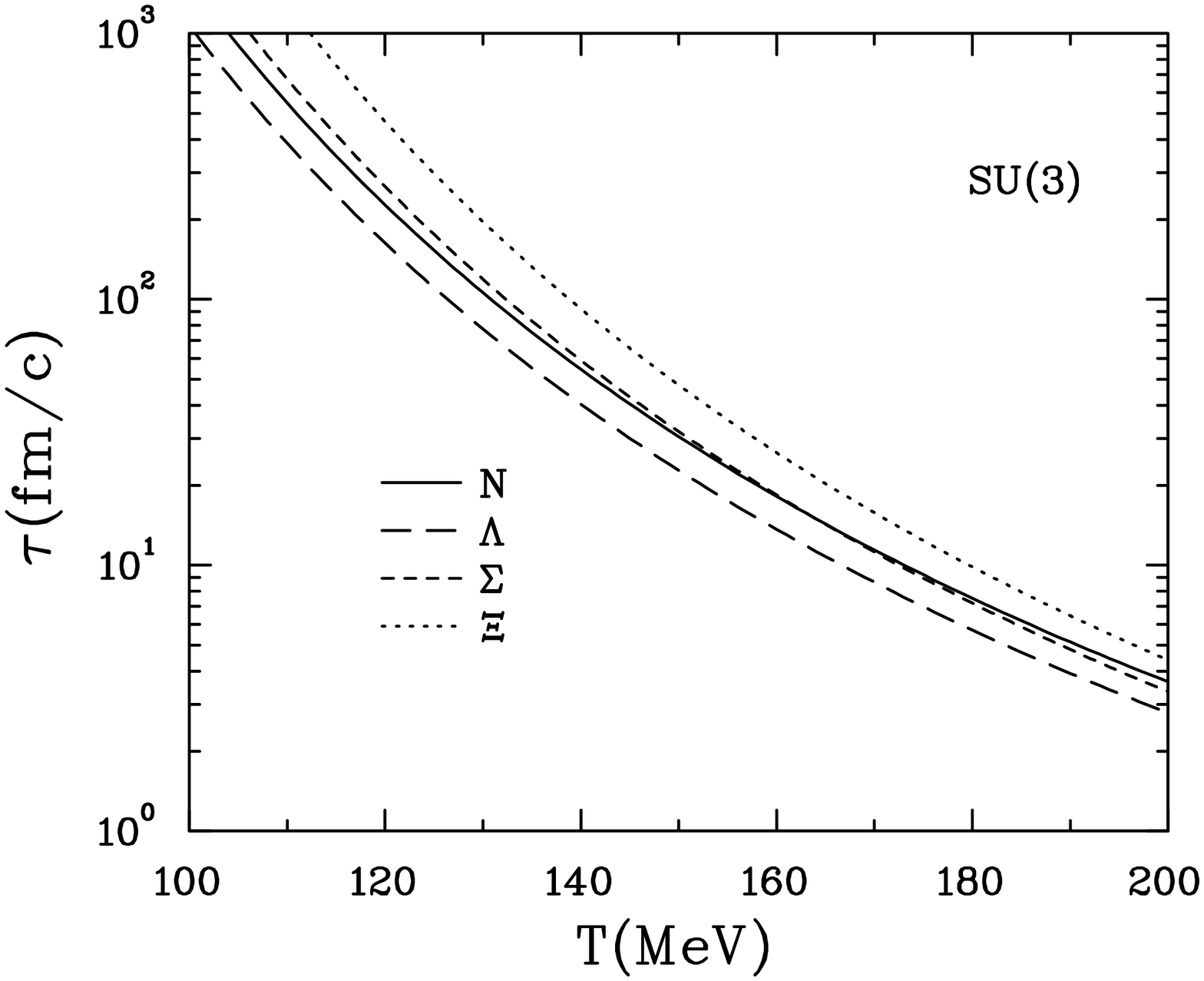,width=14cm,angle=90,bbllx=4pt,bblly=62pt,bburx=564pt,bbury=666pt}
\caption{Numerical results for equilibration times in the
case when all fluctuations in the $SU(3)$ meson sector 
are taken into accout.}
\label{fig_tau_su3}
\end{figure}

We plot this time for the proton, lambda, sigma, and cascade baryons (same
as for the anti-baryons in net baryon-free matter) in
Figs.~\ref{fig_tau_su2} and \ref{fig_tau_su3}.  Figure \ref{fig_tau_su2}
shows the times when only fluctuations in the SU(2) meson sector (no
strangeness) are allowed while Fig.~\ref{fig_tau_su3} shows the times when
all mesons or currents are included (including those with strangeness).  
The equilibration times are strongly decreasing functions of increasing
temperature.  That is typical of thermal processes; thermal rates are
generally strongly increasing functions of temperature because of the
dominant Boltzmann factor. The time for nucleons is the shortest in the
SU(2) case, which is quite natural since nucleons contain no strange
quarks and they are the lightest baryon species.  The lambda has the
longest equilibration time as a consequence of the magnitude of its
couplings to the SU(2) fluctuations.  Going to the full SU(3)  
fluctuations, see Fig.~\ref{fig_tau_su3}, shortens the cascade time
considerably but, surprisingly, the lambda now has the shortest
equilibration time! It is interesting to note that the $SU(3)$ symmetry 
is broken not only by the different masses of the baryons but also by 
the mixing of the singlet and octet mesons.

As mentioned already, to compare with data from heavy ion collisions
requires solving rate equations in an expanding and cooling system.  
Examples of how this may be done is described in \cite{KMR,KM,Greiner}.  
In addition, feed down from the decay of higher mass baryon resonances
will contribute to the observed yields.  But to get a rough idea, suppose
that the expansion time scale is about 10 fm/c and that there are full
SU(3) fluctuations in the system.  Draw a horizontal line at 10 fm/c in
Fig.~\ref{fig_tau_su3}. The intersection with the various baryon species
would suggest that these baryons would reflect a freezeout temperature in
the range 168 to 180 MeV, the exact value depending on the species. This
is approximately the range of chemical equilibration temperatures recently
seen in Au-Au collisions at 130 and 200 GeV at RHIC \cite{QM2002}.

\section{Conclusion}

In this paper we have calculated the production of spin-1/2
baryon/anti-baryon pairs through fluctuations in the strong interaction
currents.  The most basic formulation used a version of the
fluctuation-dissipation theorem that does not rely on the system being in
thermal equilibrium.  If one has a model for these fluctuations those
formulas may be used directly.  We evaluated them in thermal equilibrium,
which gives rise to equilibration times short enough that nucleons and
hyperons may very well be in chemical equilibrium in heavy ion collisions
at RHIC energies.

Two natural extensions of our work arise.  The first is to carry out the
analogous calculation for the spin-3/2 baryon decuplet.  Coupling of the
strong interaction currents to spin-3/2 baryons is much more uncertain
than the coupling to spin-1/2.  The second is to apply the formulas
derived in this paper to a dynamical model of the expanding matter.  Only
then will we be able to make direct contact with RHIC experiments.

\section*{Acknowledgements}

The authors thank Paul Ellis for useful discussions. This work was 
supported by the US Department of Energy under grant DE-FG02-87ER40328.


\newpage

\noindent
Table I: Relative strength of vector couplings for $\alpha_{V}=0$
and $\beta_{V}=1$. The names of vector mesons that couple to 
the given baryon/anti-baryon pairs are shown explicitly.\\[5mm]
\begin{tabular}{|c||c|c|c|c|c|c|c|c|}
\hline
\hspace*{10mm}& $p$  & $n$ & $\Lambda$ & $\Sigma^{0}$ & $\Sigma^{+}$ &
$\Sigma^{-}$ & $\Xi^{0}$ & $\Xi^{-}$ \\
\hline
\hline

$\overline{p}$ &
\begin{tabular}{c} $ \rho^{0}$ \\[-2mm]
                    $3 \omega$ \end{tabular}&
$\sqrt{2} \rho^{+}$ &
$-\sqrt{3} K_{*}^{+}$ &
$- K_{*}^{+}$ & $-\sqrt{2} K_{*}^{0}$ &
$0$ & $0$ & $0$ \\
\hline
$\overline{n}$ & $\sqrt{2} \rho^{-}$ &
\begin{tabular}{c} $- \rho^{0}$\\[-2mm]
                    $3 \omega$ \end{tabular} &
$-\sqrt{3} K_{*}^{0}$
& $ K_{*}^{0}$ & $0$ & $-\sqrt{2} K_{*}^{+}$ &
$0$ & $0$ \\
\hline
$\overline{\Lambda}$ & $-\sqrt{3} K_{*}^{-}$ &
$-\sqrt{3} \overline{K}_{*}^{0}$ &
\begin{tabular}{c} $2 \omega$ \\[-2mm]
                    $-\sqrt{2} \phi$ \end{tabular} &
$0$ &  $0$ & $0$  &
$\sqrt{3} K_{*}^{0}$ & $\sqrt{3} K_{*}^{+}$ \\
\hline
  $\overline{\Sigma}^{0}$  &
$- K_{*}^{-}$ & $ \overline{K}_{*}^{0}$ & $0$ &
\begin{tabular}{c} $2 \omega$ \\[-2mm]
                    $-\sqrt{2} \phi$ \end{tabular} &
$-2 \rho^{-}$ & $2 \rho^{+}$ &
$- K_{*}^{0}$ & $ K_{*}^{+}$ \\
\hline
  $\overline{\Sigma}^{+}$  & $-\sqrt{2} \overline{K}_{*}^{0}$ & $0$ &
$0$ &  $-2 \rho^{+}$ &
\begin{tabular}{c} $2 \rho^{0}$ \\[-2mm]
                    $2 \omega$ \\[-2mm]
                    $-\sqrt{2} \phi$ \end{tabular}  &
$0$  & $\sqrt{2} K_{*}^{+}$ & $0$ \\
\hline
  $\overline{\Sigma}^{-}$ & $0$ & $-\sqrt{2} K_{*}^{-}$ & $0$
& $2 \rho^{-}$ & $0$ &
\begin{tabular}{c} $2 \rho^{0}$ \\[-2mm]
                    $2 \omega$ \\[-2mm]
                    $-\sqrt{2} \phi$ \end{tabular} &
$0$ &$-\sqrt{2} K_{*}^{0}$\\
\hline
  $\overline{\Xi}^{0}$ & $0$ & $0$ & $\sqrt{3} \overline{K}_{*}^{0}$ &
$- \overline{K}_{*}^{0}$ &$\sqrt{2} K_{*}^{-}$ & $0$ &
\begin{tabular}{c} $ \rho^{0}$ \\[-2mm]
                    $ \omega$ \\[-2mm]
                    $-2\sqrt{2} \phi$ \end{tabular} &
$-\sqrt{2} \rho^{+}$ \\
\hline
  $\overline{\Xi}^{-}$  & $0$ & $0$  & $\sqrt{3} K_{*}^{-}$ &
$ K_{*}^{-}$ &$0$ & $\sqrt{2} \overline{K}_{*}^{0}$ &
$-\sqrt{2} \rho^{-}$ &
\begin{tabular}{c} $- \rho^{0}$ \\[-2mm]
                    $ \omega$ \\[-2mm]
                    $-2\sqrt{2} \phi$ \end{tabular} \\
\hline
\end{tabular}

\newpage

\noindent
Table II: Relative strength of axial-vector couplings for $\alpha_{A}=2/3$ 
and $\beta_{A}=-1/3$. The names of axial-vector mesons that couple
to the given baryon/anti-baryon pairs are shown explicitly. The notation 
$\tilde{f}_{1}$ stands for $f_{1}(1420)$.\\[5mm]
\begin{tabular}{|c||c|c|c|c|c|c|c|c|}
\hline
\hspace*{10mm}& $p$  & $n$ & $\Lambda$ & $\Sigma^{0}$ & $\Sigma^{+}$ &
$\Sigma^{-}$ & $\Xi^{0}$ & $\Xi^{-}$ \\
\hline
\hline
$\overline{p}$ &
\begin{tabular}{c} $ a_{1}^{0}$ \\
                    $\frac{1}{3} f_{1}$ \end{tabular}&
$\sqrt{2} a_{1}^{+}$ &
$-\frac{5}{3\sqrt{3}} K_{1}^{+}$ &
$\frac{1}{3} K_{1}^{+}$ & $\frac{\sqrt{2}}{3} K_{1}^{0}$ &
$0$ & $0$ & $0$ \\
\hline
$\overline{n}$ & $\sqrt{2} a_{1}^{-}$ &
\begin{tabular}{c} $- a_{1}^{0}$\\
                    $\frac{1}{3} f_{1}$ \end{tabular} &
$-\frac{5}{3\sqrt{3}} K_{1}^{0}$ &
$-\frac{1}{3} K_{1}^{0}$ & $0$ & $\frac{\sqrt{2}}{3} K_{1}^{+}$ &
$0$ & $0$ \\
\hline
$\overline{\Lambda}$ & $-\frac{5}{3\sqrt{3}} K_{1}^{-}$ &
$-\frac{5}{3\sqrt{3}} \overline{K}_{1}^{0}$ &
\begin{tabular}{c} $-\frac{2}{9} f_{1}$ \\
                    $-\frac{5\sqrt{2}}{9} \tilde{f}_{1}$ \end{tabular} &
$\frac{4}{3\sqrt{3}} a_{1}^{0}$ &  $\frac{4}{3\sqrt{3}} a_{1}^{-}$ &
$\frac{4}{3\sqrt{3}} a_{1}^{+}$ &
$\frac{1}{3\sqrt{3}} K_{1}^{0}$ & $\frac{1}{3\sqrt{3}} K_{1}^{+}$ \\
\hline
  $\overline{\Sigma}^{0}$  &
$\frac{1}{3} K_{1}^{-}$ & $-\frac{1}{3} \overline{K}_{1}^{0}$ &
$\frac{4}{3\sqrt{3}} a_{1}^{0}$ &
\begin{tabular}{c} $\frac{2}{3} f_{1}$ \\
                    $-\frac{\sqrt{2}}{3} \tilde{f}_{1}$ \end{tabular} &
$-\frac{2}{3} a_{1}^{-}$ & $\frac{2}{3} a_{1}^{+}$ &
$- K_{1}^{0}$ & $ K_{1}^{+}$ \\
\hline
  $\overline{\Sigma}^{+}$  & $\frac{\sqrt{2}}{3} \overline{K}_{1}^{0}$ &
$0$ & $\frac{4}{3\sqrt{3}} a_{1}^{+}$ &  $-\frac{2}{3} a_{1}^{+}$ &
\begin{tabular}{c} $\frac{2}{3} a_{1}^{0}$ \\
                    $\frac{2}{3} f_{1}$ \\
                    $\frac{\sqrt{2}}{3} \tilde{f}_{1}$ \end{tabular}  &
$0$  & $\sqrt{2} K_{1}^{+}$ & $0$ \\
\hline
  $\overline{\Sigma}^{-}$ & $0$ & $\frac{\sqrt{2}}{3} K_{1}^{-}$ &
$\frac{4}{3\sqrt{3}} a_{1}^{-}$
& $\frac{2}{3} a_{1}^{-}$ & $0$ &
\begin{tabular}{c} $-\frac{2}{3} a_{1}^{0}$ \\
                    $\frac{2}{3} f_{1}$ \\
                    $\frac{\sqrt{2}}{3} \tilde{f}_{1}$ \end{tabular} &
$0$ &$\sqrt{2} K_{1}^{0}$\\
\hline
  $\overline{\Xi}^{0}$ & $0$ & $0$ &
$\frac{1}{3\sqrt{3}} \overline{K}_{1}^{0}$ &
$- \overline{K}_{1}^{0}$ &$\sqrt{2} K_{1}^{-}$ & $0$ &
\begin{tabular}{c} $-\frac{1}{3} a_{1}^{0}$ \\
                    $-\frac{1}{3} f_{1}$ \\
                    $-\frac{2\sqrt{2}}{3} \tilde{f}_{1}$ \end{tabular} &
$\frac{\sqrt{2}}{3} a_{1}^{+}$ \\
\hline
  $\overline{\Xi}^{-}$  & $0$ & $0$  & $\frac{1}{3\sqrt{3}} K_{1}^{-}$ &
$ K_{1}^{-}$ &$0$ & $\sqrt{2} \overline{K}_{1}^{0}$ &
$\frac{\sqrt{2}}{3} a_{1}^{-}$ &
\begin{tabular}{c} $\frac{1}{3} a_{1}^{0}$ \\
                    $-\frac{1}{3} f_{1}$ \\
                    $-\frac{2\sqrt{2}}{3} \tilde{f}_{1}$ \end{tabular} \\
\hline
\end{tabular}

\newpage

\noindent
Table III: Numerical vector channel multipliers in the
expressions for the rates ($\alpha_{V}=0$ and
$\beta_{V}=1$).\\[5mm]
\begin{tabular}{|c||c|c|c|c|c|c|c|c|}
\hline
\hspace*{10mm}& $~~p~~$  & $~~n~~$ & $~\Lambda~~$ & $~\Sigma^{0}~$ & $~\Sigma^{+}~$ &
$~\Sigma^{-}~$ & $~\Xi^{0}~$ & $~\Xi^{-}~$ \\
\hline
\hline
$\overline{p}$ &
$2$ & $2$ & $3$ & $1$ & $2$ & $0$ & $0$ & $0$ \\
\hline
$\overline{n}$ &
$2$ & $2$ & $3$ & $1$ & $0$ & $2$ & $0$ & $0$ \\
\hline
$\overline{\Lambda}$ &
$3$ & $3$ & $\frac{8}{9}$ & $0$ &  $0$ & $0$ & $3$ & $3$ \\
\hline
   $\overline{\Sigma}^{0}$  &
$1$ & $1$ & $0$ & $\frac{8}{9}$ & $4$ & $4$ & $1$ & $1$ \\
\hline
   $\overline{\Sigma}^{+}$  &
$2$ & $0$ & $0$ & $4$ & $\frac{44}{9}$ & $0$ & $2$ & $0$ \\
\hline
   $\overline{\Sigma}^{-}$ &
$0$ & $2$ & $0$ & $4$ & $0$ & $\frac{44}{9}$ & $0$ & $2$ \\
\hline
   $\overline{\Xi}^{0}$ &
$0$ & $0$ & $3$ & $1$ & $2$ & $0$ & $\frac{26}{9}$ & $2$ \\
\hline
  $\overline{\Xi}^{-}$ &
$0$ & $0$ & $3$ & $1$ & $0$ & $2$ & $2$ & $\frac{26}{9}$ \\
\hline
\end{tabular}
\vspace{15mm}


\noindent
Table IV: Numerical axial-vector channel multipliers in the
expressions for the rates ($\alpha_{A}=2/3$ and
$\beta_{A}=-1/3$).\\[5mm]
\begin{tabular}{|c||c|c|c|c|c|c|c|c|}
\hline
\hspace*{10mm}& $~~p~~$  & $~~n~~$ & $~\Lambda~~$ & $~\Sigma^{0}~$ & $~\Sigma^{+}~$ &
$~\Sigma^{-}~$ & $~\Xi^{0}~$ & $~\Xi^{-}~$ \\
\hline
\hline
$\overline{p}$ &
$\frac{82}{81}$ & $2$ & $\frac{25}{27}$ & $\frac{1}{9}$ & $\frac{2}{9}$ & 
$0$ & $0$ & $0$ \\
\hline
$\overline{n}$ &
$2$ & $\frac{82}{81}$ & $\frac{25}{27}$ & $\frac{1}{9}$ & $0$ & 
$\frac{2}{9}$ & $0$ & $0$ \\
\hline
$\overline{\Lambda}$ &
$\frac{25}{27}$ & $\frac{25}{27}$ & $\frac{104}{729}$ & $\frac{16}{27}$ & 
$\frac{16}{27}$ & $\frac{16}{27}$ & $\frac{1}{27}$ & $\frac{1}{27}$ \\
\hline
   $\overline{\Sigma}^{0}$  &
$\frac{1}{9}$ & $\frac{1}{9}$ & $\frac{16}{27}$ & $\frac{8}{81}$ & 
$\frac{4}{9}$ & $\frac{4}{9}$ & $1$ & $1$ \\
\hline
   $\overline{\Sigma}^{+}$  &
$\frac{2}{9}$ & $0$ & $\frac{16}{27}$ & $\frac{4}{9}$ & $\frac{44}{81}$ & 
$0$ & $2$ & $0$ \\
\hline
   $\overline{\Sigma}^{-}$ &
$0$ & $\frac{2}{9}$ & $\frac{16}{27}$ & $\frac{4}{9}$ & $0$ & 
$\frac{44}{81}$ & $0$ & $2$ \\
\hline
   $\overline{\Xi}^{0}$ &
$0$ & $0$ & $\frac{1}{27}$ & $1$ & $2$ & $0$ & $\frac{26}{81}$ & 
$\frac{2}{9}$ \\
\hline
  $\overline{\Xi}^{-}$ &
$0$ & $0$ & $\frac{1}{27}$ & $1$ & $0$ & $2$ & $\frac{2}{9}$ & 
$\frac{26}{81}$ \\
\hline
\end{tabular}


\end{document}